
\vsize 8.7in

\def\doublespace{\baselineskip 22.76 pt}

\hsize 6.5 true in
\hoffset=0. true in
\voffset=0. true in
\def\mathnew{\mathsurround=0pt}
\def\simov#1#2{\lower .5pt\vbox{\baselineskip0pt \lineskip-.5pt
\ialign{$\mathnew#1\hfil##\hfil$\crcr#2\crcr\sim\crcr}}}
\def\simgreat{\mathrel{\mathpalette\simov >}}
\def\simless{\mathrel{\mathpalette\simov <}}
\font\twelverm=cmr10 scaled 1200
\font\ninerm=cmr7 scaled 1200
\font\sevenrm=cmr5 scaled 1200
\font\twelvei=cmmi10 scaled 1200
\font\ninei=cmmi7 scaled 1200
\font\seveni=cmmi5 scaled 1200
\font\twelvesy=cmsy10 scaled 1200
\font\ninesy=cmsy7 scaled 1200
\font\sevensy=cmsy5 scaled 1200
\font\twelveex=cmex10 scaled 1200
\font\twelvebf=cmbx10 scaled 1200
\font\ninebf=cmbx7 scaled 1200
\font\sevenbf=cmbx5 scaled 1200
\font\twelveit=cmti10 scaled 1200
\font\twelvesl=cmsl10 scaled 1200
\font\twelvett=cmtt10 scaled 1200
\skewchar\twelvei='177 \skewchar\ninei='177 \skewchar\seveni='177
\skewchar\twelvesy='60 \skewchar\ninesy='60 \skewchar\sevensy='60
\def\twelvepoint{\def\rm{\fam0 \twelverm}
  \textfont0=\twelverm \scriptfont0=\ninerm \scriptscriptfont0=\sevenrm
  \rm
  \textfont1=\twelvei \scriptfont1=\ninei \scriptscriptfont1=\seveni
  \def\mit{\fam1 } \def\oldstyle{\fam1 \twelvei}
  \textfont2=\twelvesy \scriptfont2=\ninesy \scriptscriptfont2=\sevensy
  \def\cal{\fam2 }
  \textfont3=\twelveex \scriptfont3=\twelveex \scriptscriptfont3=\twelveex
  \textfont\itfam=\twelveit \def\it{\fam\itfam\twelveit}
  \textfont\slfam=\twelvesl \def\sl{\fam\slfam\twelvesl}
  \textfont\bffam=\twelvebf \scriptfont\bffam=\ninebf
    \scriptscriptfont\bffam=\sevenbf \def\bf{\fam\bffam\twelvebf}
  \textfont\ttfam=\twelvett \def\tt{\fam\ttfam\twelvett}
  }
\font\simlessrgebf=cmbx10 scaled 1600
\font\twelvess=cmss10 scaled 1200
\twelvepoint
\doublespace
\def\ref{\par\noindent\hangindent=2pc \hangafter=1 }

\def\asap{{\it Astr. Ap.}}
\def\apj{{\it Ap.~J.}}
\def\apjl{{\it Ap.~J. (Letters)}}

\def\nature{{\it Nature}}
\def\cm{\hbox{cm}}
\def\s{\hbox{s}}

\def\g{\hbox{g}}

\null
\centerline{To be published in the Astrophysical Journal}
\null\vskip 0.85 true in
\centerline{\simlessrgebf A Self-Consistent Model For The Long-Term}
\vskip 0.2in
\centerline{\simlessrgebf Gamma-Ray Spectral Variability of Cygnus X-1}
\vskip 1.7in
\centerline{{\bf Fulvio Melia}\footnote{\hbox{$\null^1$}}{Presidential Young
Investigator and
Alfred P. Sloan Fellow.} and {\bf Ranjeev
Misra}\footnote{\hbox{$\null^2$}}{NASA Space Grant Fellow}}
\vskip 0.05in
\centerline{\sl Department of Physics and Steward Observatory, University of
Arizona, Tucson AZ 85721}
\centerline{\sl [melia@sgrastar.as.arizona.edu]}
\vfill\eject
\null
\centerline{\bf ABSTRACT}
\bigskip
{\twelvess
The long-term transitions of the black hole candidate Cygnus X-1 (between the
states $\gamma_1$, $\gamma_2$,
and $\gamma_3$) include the occasional appearance of a strong $\sim$ MeV bump
($\gamma_1$), whose strength appears to be anti-correlated with the continuum
flux
($\simless 400$ keV) due to the Compton upscattering of cold disk
photons by the inner, hot corona.  We develop a self-consistent disk picture
that accounts
naturally for these transitions and their corresponding spectral variations.
We argue that
the bump is due to the self-Comptonization of bremsstrahlung photons emitted
predominantly
near the plane of the corona itself.  Our results suggest that a decrease by a
factor
of $\approx 2$ in the viscosity parameter $\alpha$ is responsible for quenching
this bump and
driving the system to the $\gamma_2$ state, whereas a transition from
$\gamma_2$ to $\gamma_3$
appears to be induced by an increase of about $25\%$ in the accretion rate
$\dot M$.  In view
of the fact that most of the transitions observed in this source seem to be of
the
$\gamma_2-\gamma_3$ variety, we conclude that much of the long term gamma-ray
spectral
variability in Cygnus X-1
is due to these small fluctuations in $\dot M$.  The unusual appearance of the
$\gamma_1$
state apparently reflects a change in the dissipative processes within the
disk.
}
\bigskip\noindent {\it Subject headings}: accretion disks---black hole
physics---radiation mechanisms:
\par\noindent bremsstrahlung---radiation mechanisms:  inverse
Compton---relativity---stars: individual:  Cygnus X-1
\vfil\eject
\centerline{\bf 1. Introduction}
\medskip
The black hole candidate Cygnus X-1 has shown variability of its X-ray emission
on
all time scales ranging from milliseconds to years.  Early observations of this
strong
hard X-ray source (e.g., Tananbaum et al. 1972;  Holt et al. 1976;  Ogawara et
al. 1982)
revealed long-term transitions between two principal states:  the so-called
``low-state'' with
a relatively small ratio of soft to hard X-ray flux, and the ``high-state''
corresponding
to a softer spectrum.  Generally speaking, the hard X-ray/$\gamma$-ray emission
has been
successfully interpreted in terms of the Compton upscattering of soft disk
photons by an
optically thin, high-temperature plasma ($\sim 10^9$ K) situated either at the
inner edge of the disk (Thorne \& Price 1975;  Shapiro, Lightman \& Eardley
1976),
or within a coronal layer overlying the geometrically thin but optically thick
cooler disk
(Liang \& Price 1977;  Sunyaev \& Titarchuk 1985).  The first scenario could
arise as a
result of the secular instability present in the cool disk's inner region
(Lightman \&
Eardley 1974), which would swell the optically thick, radiation-pressure
dominated
gas to a hot, gas-pressure dominated, optically thin cloud.

More recent observations with the JPL High Resolution Gamma-Ray Spectroscopy
experiment
onboard {\it HEAO 3} (Ling et al. 1987) have suggested a more complicated
behavior than that assumed in a
simplified two-state model.  In particular, a new $\gamma$-ray emitting
state of Cygnus X-1 has been identified, in which the luminosity within an
unusually
strong $\gamma$-ray ``bump'' between $\sim 400$ keV and $\sim 2$ MeV exceeds
that of the
$50-400$ keV hard X-rays and is comparable to the overall emission below $\sim
400$ keV.
This state, labeled $\gamma_1$, together with the other low-state (dubbed
$\gamma_2$)
and the high-state ($\gamma_3$), seem to form a sequence of transitions
triggered, perhaps,
by variations in the accretion rate, as proposed earlier for the two-state
system
(e.g., Kazanas 1986).

Liang and Dermer (1988) have interpreted this $\gamma$-ray bump as due to the
emission
from a hot quasi-spherical pair-dominated cloud (distinct from the corona
producing the continuum),
in which the pair-balance condition determines the compactness parameter and
the Thomson
depth as a function of the equilibrium temperature.  They find that a
reasonable fit to the
``bump'' may be obtained for a lepton temperature $\approx 400$ keV and a pair
cloud radius of
about $9\; r_g$, where $r_g\equiv 2GM/c^2$ is the Schwarzschild radius for a
black
hole mass $M$.  Our purpose in this {\it Letter} is to build on the strengths
of these
earlier models to arrive at a self-consistent accretion disk picture that
accounts naturally for these
transitions and their corresponding spectral variations.  In particular, we
relax the assumption
that the inner hot plasma is uniform (Shapiro, Lightman \& Eardley 1976;  Liang
\& Dermer 1988) and
consider a more realistic stratified structure consistent with gravitational
settling.  As we shall
see, the secular instability still results in a two-temperature corona in the
inner region of the
disk, but the electron density in the disk plane may now be sufficiently high
to produce a copious
supply of bremsstrahlung photons.  We shall argue that the self-Comptonization
of this spectral
component is the origin of the bump seen in the $\gamma_1$ state.  Not
surprisingly, the
physical conditions in this inner hot corona are not unlike those inferred for
the pair
plasma by Liang and Dermer (1988),
but here the lepton density is due mostly to accretion, not pair production.

\bigskip
\centerline{\bf 2. A Self-Consistent Two-Temperature Disk}
\medskip
\centerline{\it 2.1. The Structure Equations}
\medskip
Our starting point is to assume a standard $\alpha$-disk model (Shakura \&
Sunyaev 1973),
which however is subject to the secular instability in the inner region that
drives
the disk from a cool state (with electron temperature $T_e\sim 10^6$ K) to the
hot,
two-fluid state ($T_e\sim 10^9$ K and ion temperature $T_i\gg T_e$;  Shapiro,
Lightman
\& Eardley 1976, hereafter SLE).  Unlike the earlier work, we do not employ
vertically-averaged equations for the coronal structure, but rather allow for
density stratification as discussed above.  For simplicity, we assume constant
temperatures in the vertical (i.e., $z$) direction, though not in the
(cylindrical) radial direction ($R$).  As such, SLE's equations (3) and (4) are
replaced by
$$
{dP_g\over dz} = -{GM\rho z\over r^3}\;\eqno(1)
$$
and
$$
\int_{-\infty}^\infty \alpha P_g(z) \,(2\pi R)\,R\,dz=(GMR)^{1/2}\dot M\,{\cal
J}\;,\eqno(2)
$$
respectively, where $r^2=R^2+z^2$, $P_g$ is the gas pressure, $\rho$ is the
density
and ${\cal J}\equiv 1-(3r_g/r)^{1/2}$.  We ignore the contribution to the
lepton
number density from pair production and check {\it a posteriori} that this is
indeed
a valid approximation.  In addition, the electron-ion energy exchange flux (SLE
equation 10) generalizes to
$$
\int dF = \int {3\over 2}\nu_E\,\rho\,k(T_i-T_e)/m_p\;dz\;,\eqno(3)
$$
where $\nu_E$ is the electron-ion coupling rate.

It is not difficult to see that with these modifications, the two-temperature
structure
of the hot inner corona is now specified by
$$
\rho=\rho_0\,\exp(-z^2/z_0^2)\;,\eqno(4)
$$
where
$$
z_0\approx (1.7\times 10^5\;\cm)\left[{{\cal J}\over \alpha}\left({M\over
3\,M_\odot}\right)
\left({\dot M\over 10^{17}\,\g\,\s^{-1}}\right)\right]^{1/3}\left({\rho_0\over
5\times 10^{-5}\,\g\,\cm^{-3}}\right)^{-1/3}\left({2r\over
r_g}\right)^{1/2}\;,\eqno(5)
$$
and the temperatures are given by
$$
T_e+T_i\approx {(8.3\times 10^{11}\;\hbox{K})\over
(2r/ r_g)^2}\left[{{\cal J}\over\alpha}
\left({\dot M\over 10^{17}\,\g\,\s^{-1}}\right)\right]^{2/3}\left({M\over
3\,M_\odot}\right)^{-4/3}
\left({\rho_0\over 5\times 10^{-5}\,\g,\cm^{-3}}\right)^{-2/3}\eqno(6)
$$
$$
{T_e\over (T_i-T_e)^{2/3}}\approx \left[{8103\over\alpha^2{\cal
J}^4}\left({M\over
3\,M_\odot}\right)^{14}\left({\dot M\over 10^{17}\,\g\,\s^{-1}}\right)^{-4}
\left({\rho_0\over 5\times 10^{-5}\,\g\,\cm^{-3}}\right)^{10}\left({2r\over
r_g}\right)^{21}\right]^{1/9}\;.\eqno(7)
$$

This prescription, however, is inappropriate whenever the physical conditions
are such that the protons attain sufficient energy to escape from the system.
Since the virial temperature
$$
T_{vir}\equiv {1\over 6}\,m_pc^2\,\left({r_g\over r}\right)\;,\eqno(8)
$$
where $m_p$ is the proton mass, decreases with increasing radius $r$, it is
anticipated
that the ion temperature $T_i$ may exceed $T_{vir}$ at large radii.
In that case, a coronal mass (and energy) outflow will prevent $T_i$ from
greatly surpassing $T_{vir}$.  For the purpose of this calculation, we shall
assume that the fraction of dissipated power lost in this wind is at most a
small fraction of the radiated luminosity.  (This restriction will be relaxed
in future radiative-hydrodynamical simulations of the accretion geometry.)
To simplify matters further, we shall also assume that in this
region (where the upper layers of the corona merge into a ``wind''),
the ion temperature is in fact equal to the virial temperature (though
of course this can only be approximately correct).  Instead of the above set of
equations, the relevant expressions that determine the coronal structure are
then
Equations (1), (3), and the condition $T_i=T_{vir}$, which yield
$$
\rho=\rho_0(1+z^2/R^2)^{-3/2}\;,\eqno(9)
$$
and
$$
T_e=5.2\times 10^9\,\hbox{K}\;\left({R\over r_g}\right)^2
\left({\rho_0\over 5\times 10^{-5}\,\g\,\cm^{-3}}\right)^{4/3}
\left({\dot M\over 10^{17}\,\g\,\s^{-1}}\right)^{-2/3}
\left({M\over3\,M_\odot}\right)^{2}\;.\eqno(10)
$$
In writing these equations, we have assumed that although the outer coronal
layers may
merge into a transonic flow (at say $z_{sonic}$), most of the structure of
interest
to us here lies at $z\ll z_{sonic}$, for which Equation (1) is still an
adequate
representation of the vertical density profile near the equatorial plane.
\medskip
\centerline{\it 2.2. Method Of Solution}
\medskip
Following SLE, we have taken the outer boundary of the two-temperature inner
region
to lie at a radius $R_0$ given by the condition $P_r(R_0)=3P_g(R_0)$, where
$P_r$ is the radiation pressure in the $\alpha$-disk.  We note, however, that
our results are insensitive
to the actual location of this boundary, as long as
$P_r(R_0)\sim\hbox{O}[P_g(R_0)]$.
For example, when $R_0$ is instead fixed by the condition
$P_r(R_0)=3/2\,P_g(R_0)$,
our spectra differ from those presented below by at most $10\%$, well within
the observational
uncertainties.
In this model, the cooling results from the inverse Comptonization of both the
cold
disk photons penetrating into the corona and the bremsstrahlung radiation
produced
in this inner hot region.  To handle this dichotomy, we divide the corona into
concentric rings and proceed as follows.  We assume
a fractional cooling by each spectral component in the first outer zone and
then iterate over the mid-plane density until the local
bremsstrahlung-self-Compton
emissivity (within this ring) is balanced by the assumed
fraction of the dissipation rate at this location.  Within any given ring,
the contribution to the overall flux from the Comptonized (outer-disk) photons
is proportional to the total number of scattering particles and the electron
temperature within
that ring, and the attenuated soft photon number density, as reflected by
the dependence of the Kompane'ets equation on these parameters.
Thus, in subsequent rings, we iterate on the local mid-plane density until the
calculated fraction
agrees with the value extrapolated on the basis of this proportionality from
the
adjacent outer ring. From this we determine the overall coronal luminosity,
including both components, and then iterate on the assumed fraction in the
outer ring
until this power matches the total dissipation rate
inside the corona. We check ring by ring whether the first set of structure
equations (see \S 2.1. above) yields an ion temperature $T_i>T_{vir}$.  If mass
loss
is indicated at a particular radius, we then use the second set of equations to
determine the local structure.

The cooling rate
(and concomitantly the spectrum) is obtained by solving the relativistic
steady-state
Fokker-Planck equation, whose diffusion coefficient is valid for {\it
arbitrary} photon
and electron energies (Prasad et al. 1988).  At the temperatures of interest in
our
problem, the use of the non-relativistic Kompane'ets equation (Kompane'ets
1956) is
not valid, which in the past has led to the development of a (time-consuming)
Monte-Carlo
scheme in order to treat the inverse Compton scattering correctly (e.g., Liang
\&
Dermer 1988).  With the recent development of the exact analytical formula for
this (relativistically-correct) diffusion coefficient of the Compton
Fokker-Planck equation,
we are now able to circumvent this difficulty by simply solving the
relativistic Kompane'ets
equation without any approximation.  In this regard, our approach differs from
that of SLE in at least 2 ways.  First, our source term includes both the cool
disk photons penetrating the $y\simless 1$ region of the corona and the
bremsstrahlung photons produced within the corona itself (predominantly near
the plane).
Here, $y\equiv (4kT_e/m_ec^2)\hbox{Max}(\tau,\tau^2)$ is the dimensionless
parameter that characterizes Comptonization, and $\tau$ is the scattering
optical depth.  Secondly, since our equation is valid at all energies,
our results are not subject to the restrictions (e.g., in $T_e$) and errors
inherent
in Cooper's modification to the Kompane'ets equation (Cooper 1971).  Of course,
the main result of our calculations is the spectrum, and this is determined
along the lines developed by SLE, in which the energy-dependent rate of photon
loss appearing in the Fokker-Planck equation also acts as the source
term for the energy-dependent photon flux.

\centerline{\bf 3. The Gamma-Ray Spectral Variability}
\medskip

Since the structure of the hot inner corona is here calculated
self-consistently,
we have at most only 3 physical quantities that need to be fixed.  These are
the black hole mass
$M$, the accretion rate $\dot M$, and the viscosity parameter $\alpha$.  A very
conservative lower mass limit for Cygnus X-1 is $3.4\,M_\odot$ (e.g.,
Paczy\'nski
1974), whereas the most likely value is thought to lie in the range
$9-15\;M_\odot$
(e.g., Avni \& Bahcall 1975).  We have chosen to work with a value
$M=10\;M_\odot$,
though again our results are insensitive to the actual mass as long as
$5\;M_\odot
\simless M\simless 15\;M_\odot$.  The basis for this ``robustness'' is the fact
that
distances scale as the Schwarzschild radius $r_g=2GM/c^2$, so that the overall
energy budget ($\sim GM\dot M/r$) is independent of $M$ (to first order) for a
given
$\dot M$.  Nonetheless, the coronal structure does change with $M$, but we have
found
only $\simless 10\%$ differences in the spectra over this mass range and have
therefore
not considered $M$ to be a principal parameter characterizing our solutions.
In
addition, although we expect (slight) variations in $\dot M$ between the
various
states of Cygnus X-1, the gross X-ray/$\gamma$-ray energetics for an assumed
source distance of
$2.5$ kpc restrict the accretion rate to a value $\approx 10^{18}$ g s$^{-1}$.

Our solutions for the $\gamma_1$, $\gamma_2$, and $\gamma_3$ states are
compared to the
data in Figures 1, 2, and 3.  The data used by SLE overlap most significantly
with
the $\gamma_1$ spectrum and have therefore been included in Figure 1.
We note also that we have excluded the 1.5 MeV point in the HEAO 3 data since
it may be
contaminated by a background line (Liang and Dermer 1988).  In each case, the
dashed curve shows the contribution from the bremsstrahlung/self-Comptonized
(BSC) photons originating
from within the corona itself, whereas the solid curve represents the overall
spectrum comprising
the BSC radiation, the cold disk photons that are Compton upscattered in the
$y\simless 1$ region
of the corona, and the thermal emission from the disk itself (which dominates
at $E\simless 3$ keV).
What emerges from these results is the interesting correlation between the
variation in $\alpha$
and the gradual
shift of emitted power from the $\gamma$-ray continuum into the broad $\sim$
MeV bump (see Figure 4).
In particular, a factor $2$ decrease in $\alpha$ is sufficient to drive the
system from the $\gamma_1$
to the $\gamma_2$ states.  Thus, the emergence of the
bump in the $\gamma_2$ and especially the $\gamma_1$ states must be accompanied
by a corresponding
reduction in the continuum flux below about $400$ keV, as confirmed by the
data.
On the other hand, an increase in $\dot M$ raises the flux in both the
continuum and
the bump.  It appears that an increase of about $25\%$ in $\dot M$ is
responsible for driving
the system from the $\gamma_2$ to the $\gamma_3$ states (see Figure 5).

Figures 6, 7, 8, and 9 demonstrate how the internal structure of the corona
changes in response to a
variation in either $\alpha$ (characterizing the transition from the $\gamma_1$
to
the $\gamma_2$ states) or the accretion rate (which induces the transition from
$\gamma_2$ to
$\gamma_3$).  In Figure 6, the most significant aspect of the electron
temperature profile
is that the corona becomes optically thick toward small radii due to
gravitational stratification
(see the profiles of mid-plane density shown in Figure 7), and $T_e$
increases rapidly in this region to enhance the bremsstrahlung-self-Compton
cooling process.
The profiles of ion temperature are shown in Figure 8.  In all three states,
$T_i$ increases
with radius until it reaches the virial temperature $T_{vir}$, though the
radius at which this
occurs is significantly larger in the $\gamma_1$ state than the other two.
The impact of this change in the ion temperature is reflected in the profiles
of the $y$-parameter
shown in Figure 9.  Here, the solid curves correspond to the location in the
corona where
$y=1$, whereas the dashed curves are for $y=0.5$.  As one would expect, the
corona
appears more extended in those regions where $T_i\approx T_{vir}$.  The
temperature, $T_e$
jumps to a value $\approx 50$ keV near the boundary (at $R_0\approx 15 r_g$)
where the instability first
arises, and then continues to increase towards $\approx 350$ keV as the matter
approaches $\sim 6-7 r_g$.
Thus, since most of the bremsstrahlung photons are emitted at $y>1$, it is
evident from Figure 5 that the location
of the bump at $\sim 1$ MeV is a natural consequence of the self-Comptonization
of the radiation, which
approaches a Wien peak centered at roughly $3 kT_e$ in this inner hot region.
However, the midplane density
$\rho_0$ is lower in the $\gamma_2$ and $\gamma_3$ states compared to
$\gamma_1$, and since
the  bremsstrahlung intensity scales as $\rho_0^2$, the appearance of the bump
is more evident in $\gamma_1$
than $\gamma_2$ and $\gamma_3$.

Using these results, we may now justify {\it a posteriori} our neglect of pairs
in the equilibrium
lepton number density.  Under the simplifying (and conservative) assumption
that the
radiation distribution within the corona is isotropic, we estimate the pair
production rate via $\gamma\gamma$ interactions
to be about $9\times 10^{40}$ s$^{-1}$ in the $\gamma_1$ state.  (The pair
creation rate is smaller in the
$\gamma_2$ and $\gamma_3$ states due to the relatively lower number density of
$E> 511$ keV $\gamma$-rays.)
The average (accretion) electron number density is $\langle n_e\rangle\approx
5\times 10^{17}$ cm$^{-3}$,
for which the positron (in-flight) annihilation time is $t_a\equiv
1/(n_e\,\sigma_T\,c)\approx 1\times 10^{-4}$
s.  This is in fact the shortest time scale associated with the positron
distribution,
providing an upper limit to the positron number density $n_{pairs}$.  We
therefore expect the
positron average equilibrium number density to be
$\langle n_{pairs}\rangle \simless 8\times 10^{13}$ cm$^{-3}$ for the coronal
structure depicted in Figure 5, in
which the scale size is $\approx 3\times 10^7$ cm.  As such, $n_{pairs}\ll n_e$
and our approach would appear
to be valid, at least in the case of Cygnus X-1.
Incidentally, if we further assume that the subsequent conversion of these
pairs into $511$ keV photons is an efficient one, the $511$ keV luminosity
should be $\approx 2.8\times
10^{36}$ ergs s$^{-1}$, or about 10 times weaker than the luminosity in the
bump.

This work was supported in part by NSF grant
PHY 88-57218, NASA grant NAGW-2380, and the Alfred P. Sloan Foundation.
We are grateful to Edison Liang for a helpful suggestion, and
to the anonymous referee for his thoughtful comments and
for pointing out an error in the original presentation of our spectra.

\vfill\eject\null
\centerline{\bf {References}}
\medskip
\noindent
\ref Avni, Y. \& Bahcall, J.N. 1975, \apjl, {\bf 202}, L131.
\ref Cooper, G. 1971, {\it Phys. Rev. D}, {\bf 3}, 2312.
\ref Holt, S.S., Boldt, E.A., Serlemitsos, P.J. \& Kaluzienski, L.J. 1976,
\apjl,
{\bf 203}, L63.
\ref Kazanas, D. 1986, \asap, {\bf 166}, L19.
\ref Kompane'ets, A.S. 1956, {\it Zh. Eksp. Teor. Fiz.}, {\bf 31}, 876.
(English transl. in {\it Soviet Phys.---JETP}, {\bf 4}, 730 [1957]).
\ref Liang, E.P. \& Dermer, C.D. 1988, \apjl, {\bf 325}, L39.
\ref Liang, E.P.T. \& Price, R.H. 1977, \apj, {\bf 218}, 247.
\ref Lightman, A.P. \& Eardley, D.M. 1974, \apjl, {\bf 187}, L1.
\ref Ling, J.C., Mahoney, W.A., Wheaton, Wm.A. \& Jacobson, A.S. 1987, \apjl,
{\bf 321}, L117.
\ref Ogawara, Y., Midsuda, K., Masai, K., Vallerga, J.V., Cominsky, L.R.,
Grunsfeld, J.M.,
Kruper, J.S. \& Ricker, G.R. 1982, \nature, {\bf 295}, 675.
\ref Paczy\'nski, B. 1974, \asap, {\bf 34}, 161.
\ref Prasad, M.K., Shestakov, A.I., Kershaw, D.S. \& Zimmerman, G.B. 1988, {\it
J. Quant.
Spectrosc. Radiat. Transfer}, {\bf 40}, 29.
\ref Schreier, E., Gursky, H., Kellogg, E. Tananbaum, H., and Giacconi, R.
1971, \apj, {\bf 170}, L21.
\ref Shakura, N.I. \& Sunyaev, R.A. 1973, \asap, {\bf 24}, 337.
\ref Shapiro, S.L., Lightman, A.P. \& Eardley, D.M. 1976, \apj, {\bf 204}, 187
(SLE).
\ref Sunyaev, R.A. \& Titarchuk, L.G. 1985, \asap, {\bf 143}, 374.
\ref Tananbaum, H., Gursky, H., Kellogg, E., Giacconi, R. \& Jones, C. 1972,
\apjl,
{\bf 177}, L5.
\ref Thorne, K.S. \& Price, R.H. 1975, \apjl, {\bf 195}, L101.
\null
\vfill\eject\null
\centerline{\bf Figure Captions}
\noindent{\bf Figure 1. }  Theoretical spectrum corresponding to the $\gamma_1$
state of
Cygnus X-1. {\it Dashed curve}: bremsstrahlung/self-Compton component (BSC);
{\it Solid curve:}
overall spectrum, including the BSC component, the disk radiation upscattered
within the
$y\simless 1$ region of the inner corona, and the thermal disk radiation at
$\simless 3$ keV.
Data for $E\simgreat 50$ keV are taken from Ling et al. 1987.  Data below $\sim
50$ keV (which however
were collected at a different epoc) are from Schreier et al. 1971.
\vskip 0.2in
\noindent{\bf Figure 2. }  Same as Figure 1, except now for the $\gamma_2$
state of Cygnus X-1.
All data are from Ling et al. 1987.
\vskip 0.2in
\noindent{\bf Figure 3. }  Same as Figure 1, except now for the $\gamma_3$
state of Cygnus X-1.
All data are from Ling et al. 1987.
\vskip 0.2in
\noindent {\bf Figure 4.}  Comparison of the $\gamma_1$
and $\gamma_2$ theoretical spectra, showing the cross-over at
approximately $400$ keV, and the anti-correlation between the strength of the
bump at $\approx
1$ MeV and the continuum flux below this cross-over energy.
\vskip 0.2in
\noindent {\bf Figure 5.}  Comparison of the $\gamma_2$
and $\gamma_3$ theoretical spectra, showing the overall shift in flux, for both
the continuum and the bump, when $\dot M$ changes by $\approx 25\%$.
\vskip 0.2in
\noindent {\bf Figure 6.}
The run of electron temperature $T_e$ as a function of
the radius $r$ in the disk.  The onset of the secular instability at $r\approx
15-17\;r_g$
is signaled by a rapid increase in both $T_e$ (to $\simgreat 50$ keV),
as specified by Equation (6) or (10).  Because of gravitational stratification,
the corona becomes
optically thick toward smaller radii, and $T_e$ increases further to enhance
the cooling rate due
to bremsstrahlung-self-Compton emission in this region.
\vskip 0.2in
\noindent {\bf Figure 7.} Profiles of the mid-plane density as a function of
radius for
each of the three gamma-ray states.  The slight jump in density reflects a
transition from
the bound corona (for which $T_i<T_{vir}$) to a ``coronal wind'' zone (in which
$T_i\approx T_{vir}$).
\vskip 0.2in
\noindent {\bf Figure 8.} The run of ion temperature $T_i$ as a function of
radius for each
of the three gamma-ray states.  $T_i$ is limited from above by the value of the
virial
temperature $T_{vir}$.
\vskip 0.2in
\noindent {\bf Figure 9.}  Contours of the Compton $y$-parameter for $y= 0.5$
(doted) and $1.0$
(solid).  (a) $\gamma_1$ state, (b) $\gamma_2$ state, and (c) $\gamma_3$ state.
Distances are in units of the Schwarzschild radius $r_g\equiv 2GM/c^2$.
\vfill\end